\documentclass[runningheads]{llncs}
\usepackage[T1]{fontenc}
\usepackage{amsmath}
\usepackage{amssymb}
\usepackage{graphicx}
\begin{document}
\title{PuriGen: Purity-Aware Deep Generative Modeling for Predicting Stem Cell Lineage Fate}
\titlerunning{PuriGen: Purity-Aware Deep Generative Modeling}
%
\author{Yanli Li, Tianying Sheng, Hala Zreiqat, ZuFu Lu, Zhiyong Wang}
\authorrunning{Y. Li et al.}
%
\institute{The University of Sydney}
\maketitle              
\begin{abstract}
Predicting stem-cell lineage fate is central to elucidating biomaterial–stem cell interactions and optimizing strategies for tissue regeneration. While existing approaches demonstrate the feasibility of lineage prediction, they remain limited in modelling compositional heterogeneity and time-dependent differentiation dynamics. To mitigate this gap, we propose PuriGen, a purity-aware deep generative framework for biomaterial-induced lineage prediction from bulk transcriptomic data. PuriGen consists of two key modules, PuritySCVI and PuritySCANVI, which extend scVI and scANVI by incorporating a pretrained GBMPurity head to provide a sample-level purity-like compositional signal. In addition, purity-aware regularisation and temporal constraints are introduced to encourage informative purity prediction and biologically consistent latent representations. Experiments on bulk mesenchymal stem-cell datasets collected across multiple biomaterial conditions and induction stages show that the proposed framework achieves better performance than the existing methods.

\end{abstract}
\section{Introduction}
Stem cells can self-renew and differentiate into various specialized lineages, allowing the regeneration and repair of tissues in multicellular organisms \cite{wang2024mitochondrial}. Over the past decades, researchers have devoted substantial effort to understanding these processes in order to support the development of regenerative medicine and tissue engineering. \cite{murry2008differentiation}. Among various stem cell types, mesenchymal stem cells (MSCs) have attracted particular interest due to their accessibility, multipotency, and immunomodulatory properties. Specifically, the fate of MSCs can be influenced by biochemical induction, physical and mechanical cues from their surrounding microenvironment \cite{chen2016fate}. Biomaterials with tunable surface structure, stiffness, and chemical composition serve as important sources of these cues. Thus, MSCs can be directed toward different lineages by designing these biomaterials, which enables applications in the tissue regeneration.

Accurately predicting stem-cell lineage fate on biomaterials helps speed up biomaterials screening and reduces experimental cost. Traditional techniques such as immunostaining and histology are effective, but they are time-consuming and typically capture only late-stage differentiation outcomes. As a result, machine learning has been applied to achieve automated analysis on how cells respond to biomaterial cues \cite{zhou2023assessing,liu2026destinynet}. Bulk RNA-sequencing (RNA-seq) and single-cell RNA-sequencing (scRNA-seq) are the main transcriptomic techniques used for this purpose.

Thanks to its stable, high-coverage, and cost-effective transcriptomic measurements, RNA-seq is well suited for large-scale biomaterial studies \cite{thind2021demystifying}. However, because RNA-seq averages gene expression across many cells with different cell types, it obscures cellular heterogeneity during differentiation in mixed populations of progenitor and partially committed cells. Such compositional heterogeneity can influence observed expression patterns and may compromise predictive accuracy and biological interpretability. In contrast, scRNA-seq captures fine-grained transcriptional dynamics and developmental trajectories at single-cell resolution, but remains less practical for large-scale screening because of its cost, technical complexity, and data sparsity \cite{stegle2015computational}. Therefore, a key research objective is to integrate their strengths and better capture heterogeneity while retaining the practical advantages of bulk RNA-seq data.

Earlier studies have shown that bulk transcriptomes can be used to predict lineage fate with traditional machine learning methods \cite{zaitsev2022precise}. For example, the MeD-P framework \cite{zhou2023assessing} achieved over 90\% accuracy using classifiers such as support vector machines, random forests, and k-nearest neighbours. However, MeD-P is purely discriminative: it learns decision boundaries without explicitly modelling the underlying biological variability, and it treats each bulk sample as homogeneous despite its mixed cellular composition \cite{liu2020machine}. This limitation has motivated the introduction of deep generative models, which can better capture biological variation and technical noise through latent representations. In particular, scVI \cite{lopez2018deep} and its semi-supervised extension scANVI \cite{xu2021probabilistic} have proven effective for learning probabilistic embeddings from transcriptomic data. Nevertheless, these models were originally designed for single-cell analysis and, when directly applied to bulk RNA-seq, they still fail to account for within-sample compositional heterogeneity and the proportion of differentiated cells, referred to here as purity, which describes the fraction of committed cells in a mixed population and strongly influences bulk gene expression.

To address this gap and further enable the trained model to leverage purity information, we propose PuriGen, a purity-aware deep generative framework. Our framework introduces purity estimation and models it as a continuous latent variable connecting gene expression, biomaterial conditions, and differentiation stages. Specifically, PuriGen consists of two designed modules: (i) PuritySCVI module extends scVI for unsupervised generative modelling of bulk RNA-seq data and incorporates a pretrained GBMPurity head to provide compositional guidance. It further introduces three additional regularisation terms: a variance-centrality term to preserve purity diversity and avoid collapse, a temporal smoothness term to model gradual differentiation, and a weak monotonicity constraint to encourage biologically plausible purity progression. (ii) PuritySCANVI module, extends scANVI for semi-supervised training, adding lineage labels and a purity-guided auxiliary Kullback–Leibler term linking classification certainty with purity confidence. Together, PuriGen can reconstruct gene-expression distributions, estimates sample purity, and predicts lineage fate within the same probabilistic structure. 

The main contributions of this study include:
\begin{itemize}
    \item We propose PuriGen, a unified purity-aware deep generative framework for stem-cell lineage prediction from bulk RNA-seq data by taking the heterogeneity nature of bulk sequence data into account in terms of purity.
    
    \item We introduce PuritySCVI, a purity-aware extension of scVI for bulk RNA-seq data, which incorporates purity estimation into the latent space and enables joint modelling of biological variation and compositional heterogeneity.

    \item We introduce PuritySCANVI, a purity-aware extension of scANVI that incorporates lineage supervision and purity guidance to improve lineage prediction performance.

    
    \item Experimental results on bulk mesenchymal stem-cell datasets collected across multiple biomaterial conditions and induction stages show that the proposed framework achieves better performance than the existing methods. 
\end{itemize}

\section{Related Work}
In this section, we review related work on deep generative models for transcriptomics and purity estimation techniques across single-cell and bulk data to provide the background for this study.

\subsection{Deep generative models for transcriptomics}

Deep generative models provide a probabilistic framework for analysing high-dimensional transcriptomic data by learning latent representations that capture biological variation while accounting for technical noise. Today, Variational autoencoders (VAEs) \cite{kingma2013auto} are widely used for this purpose and have become an important foundation for transcriptomic representation learning. 

Among these methods, scVI \cite{lopez2018deep} is a representative model that adapts VAEs to RNA-seq data using a negative-binomial likelihood and latent biological variables. Based on ScVI, scANVI \cite{xu2021probabilistic} further incorporates label information to support probabilistic annotation and class-aware representation learning. Furthermore, related models such as TOTALVI \cite{gayoso2021joint} and PEAKVI \cite{ashuach2022peakvi} show that this framework can be extended to multimodal and other omics settings.

Despite their strong performance, most existing deep generative models were developed for single-cell or multimodal data rather than compositionally mixed bulk RNA-seq samples. Related deep-learning methods such as GBMPurity \cite{thomas2025gbmpurity} have demonstrated the feasibility of predicting purity from bulk transcriptomic data. Despite this progress, most existing approaches focus on purity estimation as an isolated objective and do not integrate transcriptomic variation, sample composition, and downstream prediction into a unified probabilistic framework. 

\subsection{Purity estimation from single-cell to bulk data}

Purity estimation and deconvolution methods aim to infer the composition of bulk RNA-seq samples, which often contain mixtures of different cell states or cell types. Classical methods such as CIBERSORTx \cite{newman2019determining}, MuSiC \cite{wang2019bulk}, and BSEQ-sc \cite{baron2016single} estimate cell-type proportions from bulk transcriptomic data using reference profiles, while xCell \cite{aran2017xcell} provides enrichment-based estimates without requiring exact reference matrices. These approaches have shown that modelling sample composition can improve the interpretation of bulk transcriptomic data, but they are usually applied as separate preprocessing steps rather than being integrated into downstream predictive models. 

Recent studies have also explored deep learning for purity prediction. For example, GBMPurity \cite{thomas2025gbmpurity} uses simulated pseudobulk data to estimate tumour purity from bulk RNA-seq, showing that composition-related signals can be learned through neural models. At the same time, a growing body of work \cite{lotfollahi2019scgen} has investigated cross-scale transfer from single-cell to bulk data, using pseudobulk construction, shared latent spaces, or transfer learning to bridge the two modalities. These studies suggest that single-cell-informed representations can improve bulk-level inference. However, most existing methods still treat purity estimation, deconvolution, and downstream prediction as separate tasks, rather than jointly modelling them within a unified probabilistic framework.


\section{Purity-Aware Deep Generative Modeling Framework (PuriGen)}
\begin{figure}
    \centering
    \includegraphics[width=1.0\linewidth]{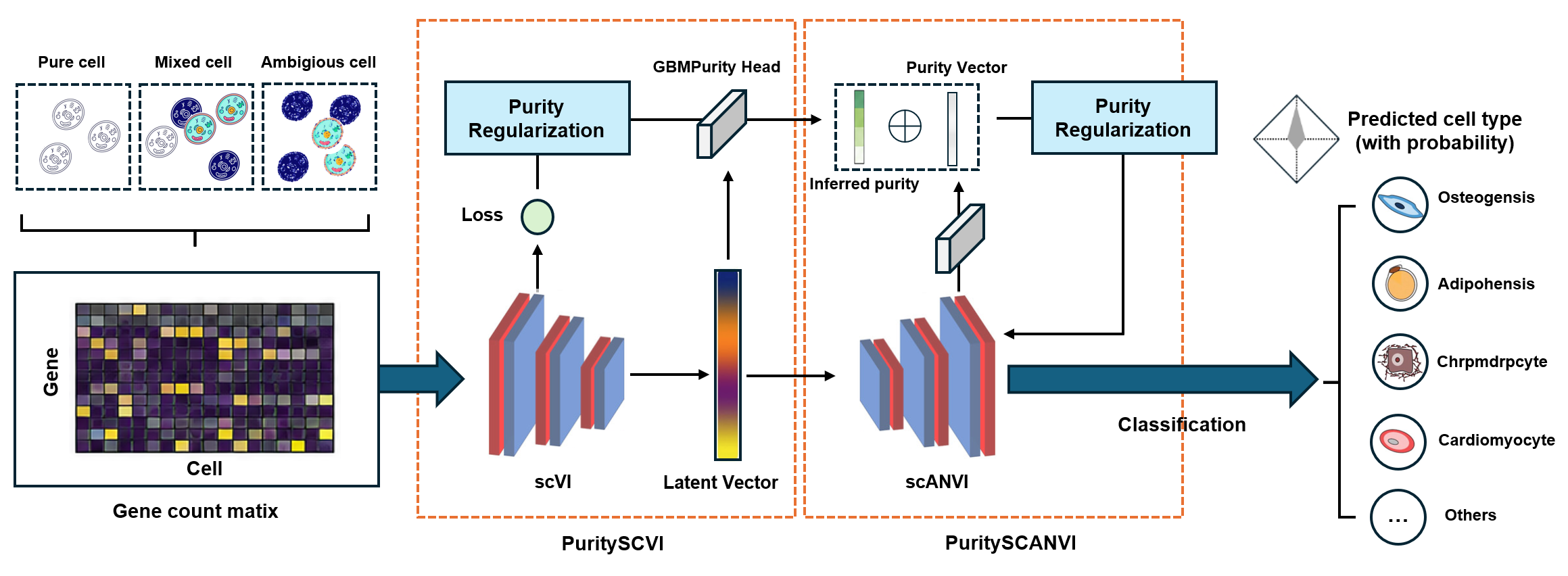}
    \caption{
Overview of PuriGen for bulk RNA-seq lineage fate prediction.
(Middle left) PurityscVI learns latent representations and estimates the purity with regularization. 
(Middle right) PurityscANVI integrates latent features, purity vectors, and partial labels for purity-guided classification and differentiation-state interpretation.
}
    \label{fig:arc}
\end{figure}

To predict biomaterial-induced stem-cell lineage fate from bulk RNA sequencing data while estimating the compositional purity of each sample, we propose a purity-aware deep generative framework (PuriGen) in this section. The full framework consists of two modules: (i) PuritySCVI, a purity-aware extension of scVI for bulk RNA-seq data that incorporates purity estimation; and (ii) PuritySCANVI, which further extends scANVI by introducing lineage supervision and purity guidance. Fig. \ref{fig:arc} illustrates the architecture of the whole framework.

\subsection{PuriGen Architecture}

\subsubsection{PuritySCVI}
PuritySCVI is responsible for the unsupervised latent–purity learning stage, learning latent representations and purity information. The loss function of PuritySCVI consists of four parts: the ELBO term, the purity regularization term, the temporal smoothness term, and the monotonicity term. Formally, the loss function of PuritySCVI is shown as followed:
\begin{equation}
\mathcal{L}_{\text{PuritySCVI}}
=
\mathcal{L}_{\text{ELBO}}
+
\mathcal{L}_{\text{pur}}
+
\mathcal{L}_{\text{smooth}}
+
\mathcal{L}_{\text{mono}}.
\end{equation}

The first term, $\mathcal{L}_{\text{ELBO}}$, is the standard evidence lower bound inherited from scVI:
\begin{equation}
\mathcal{L}_{\text{ELBO}}
= \mathbb{E}_{q_{\varepsilon}(z_n \mid x_n)}
\big[
\log p_{\omega}(x_n \mid z_n)
\big]
-
D_{\mathrm{KL}}
\big(
q_{\varepsilon}(z_n \mid x_n)
\,\|\, 
p(z_n)
\big),
\end{equation}
where $x_n$ denotes the observed bulk RNA-seq expression of sample $n$, $z_n$ is the latent variable, $q_{\varepsilon}(z_n \mid x_n)$ is the encoder posterior, and $p_{\omega}(x_n \mid z_n)$ is the decoder likelihood. This term ensures that the learned latent representation preserves the information required to reconstruct the observed gene expression while keeping the latent space regularised.

To explicitly model purity, the purity regularisation term is introduced as
\begin{equation}
\mathcal{L}_{\text{pur}}
=
\vartheta_p
\left[
-\varrho_{\mathrm{var}} \, \mathrm{Var}(\hat{p})
+
\varrho_{\mathrm{ctr}}
\frac{1}{N}
\sum_{n=1}^{N}
(\hat{p}_n - \mu_p)^2
\right],
\end{equation}
where $\hat{p}_n \in [0,1]$ denotes the predicted purity of sample $n$, $N$ is the mini-batch size, $\mu_p = 0.5$, $\varrho_{\mathrm{var}} = 0.1$, and $\varrho_{\mathrm{ctr}} = 0.02$. The variance term encourages diversity among purity predictions so that they do not collapse to a trivial constant, while the centrality term prevents the purity estimates from drifting to unreasonable extremes.

To enforce temporal consistency, the smoothness term is defined as
\begin{equation}
\mathcal{L}_{\text{smooth}}
=
\vartheta_s
\frac{1}{N_t - 1}
\sum_{i=1}^{N_t-1}
\exp\!\left(-\frac{t_i}{\varsigma}\right)
\left(
\hat{p}_{t_{i+1}} - \hat{p}_{t_i}
\right)^2,
\qquad
\varsigma = 48\,\mathrm{h},
\end{equation}
where $N_t$ is the number of ordered time points and $\hat{p}_{t_i}$ is the predicted purity at time point $t_i$. This term penalises abrupt changes in purity between neighbouring time points, with larger weights assigned to earlier stages of differentiation.

Finally, a monotonicity constraint is added to reflect the assumption that differentiation purity should generally increase over time:
\begin{equation}
\mathcal{L}_{\text{mono}}
=
\vartheta_m
\frac{1}{N_t - 1}
\sum_{i=1}^{N_t-1}
\max
\left(
0,\;
-\big(
\hat{p}_{t_{i+1}} - \hat{p}_{t_i}
\big)
\right).
\end{equation}
This term penalises decreases in purity along the temporal trajectory, while still allowing minor local variation when the decrease is small.

Taken together, these four terms enable PuritySCVI to learn a latent representation that not only reconstructs the input bulk RNA-seq data, but also captures purity-related heterogeneity and biologically plausible differentiation dynamics.

\subsubsection{PuritySCANVI} PuritySCANVI extends scANVI with purity guidance. We initialise PuritySCANVI from PuritySCVI to enable the latent space encodes compositional information. The GBMPurity head remains frozen and serves as a reference during training.

At each mini-batch, two purity signals are combined: $p_{\mathrm{scvi}}$, the purity estimated in the unsupervised stage and stored in the data object, and $\hat{p}_{\mathrm{GBM}}$, the output of the frozen GBMPurity head. A convex combination is used:
\begin{equation}
p_{\mathrm{comb}}
=
\varpi \, p_{\mathrm{scvi}}
+
(1-\varpi)\,\hat{p}_{\mathrm{GBM}},
\qquad
\varpi = 0.4.
\end{equation}

Here, $\varpi$ is a trade-off hyperparameter and $0.4$ is selected in this study based on the experimental results. The combined purity $p_{\mathrm{comb}}$ is then used to drive three terms: the purity regularisation term defined above, the temporal smoothness term with a simple curriculum weighting, and an auxiliary Kullback--Leibler divergence term that links classification confidence to purity confidence:
\begin{equation}
\mathcal{L}_{\mathrm{aux}}
=
\vartheta_a \,
\mathbb{E}
\left[
D_{\mathrm{KL}}
\bigl(
q_{\varepsilon}(y \mid x)
\,\|\, 
q^{*}_{\varepsilon}(y \mid x)
\bigr)
\cdot
w(p_{\mathrm{comb}}, t)
\right].
\end{equation}

Accordingly, the full objective of PuritySCANVI is defined as
\begin{equation}
\mathcal{L}_{\mathrm{PuritySCANVI}}
=
\mathcal{L}_{\mathrm{scANVI}}
+
\mathcal{L}_{\mathrm{pur}}
+
\mathcal{L}_{\mathrm{smooth}}
+
\mathcal{L}_{\mathrm{aux}}.
\end{equation}

\subsection{Training Pipeline}

The whole training pipeline begins with data collection and preprocessing to ensure that the data are properly prepared for subsequent modeling. Once the data is processed, PuritySCVI is trained in an unsupervised manner to learn the latent structure of bulk RNA-seq samples and to infer sample purity.  In this training phase, the model optimises the ELBO together with purity-aware regularisation terms. These terms encourage the latent space to encode not only transcriptomic variation, but also purity-related and temporal information. In particular, purity variance and centrality constraints are used to avoid trivial purity predictions, while temporal smoothness and monotonicity encourage biologically plausible purity trajectories.

Once the PuritySCVI training is finalized, the PuritySCANVI will be subsequently initialised from PuritySCVI and further trained with lineage labels, while purity information is used to guide classification. During the second training phase, the pretrained latent representation is refined for semi-supervised lineage prediction. The GBMPurity head remains frozen and serves as a reference. Its output is combined with the purity estimates learned in the first phase to construct a composite purity signal, which is then used in the purity regularisation, temporal smoothness, and auxiliary KL terms. This design allows purity information to regularise classification and improves robustness to ambiguous samples.

\section{Experiments}
In this section, we conduct the experiments to demonstrate the effectiveness of proposed purity-aware deep generative framework.

\subsection{Data Collection and Preprocessing}

The experiments were conducted on the bulk RNA-seq dataset derived from study \cite{zhou2023assessing}, which profiled human mesenchymal stem cells (hMSCs) cultured under multiple biomaterial conditions. The dataset covers three major differentiation lineages, including osteogenic, chondrogenic, and adipogenic, together with undifferentiated samples. In addition, time-series measurements were provided to capture the transcriptional progression of differentiation under different induction conditions.

We leverage the available processed expression matrices in study \cite{zhou2023assessing}, while reproducing the preprocessing workflow. Specifically, the raw reads were first cleaned to remove adapter sequences and low-quality bases. This quality-control step eliminates sequencing artefacts and preserves only high-confidence reads for alignment. Ribosomal RNA (rRNA) contamination was then removed using “bowtie2”, which aligned the cleaned reads to the Homo sapiens reference genome. Reads mapping to rRNA regions were filtered out, ensuring that only messenger-RNA (mRNA) transcripts were retained. 

The filtered reads were then aligned to the human reference genome using the STAR aligner, with genome indices generated from the GRCh38 reference sequence and its corresponding gene annotation. Gene-level quantification was then performed using featureCounts, which summarises read counts per gene based on exon annotations. The resulting count matrices represent the transcriptional abundance of each gene and form the foundation for downstream expression analysis.

Following quantification, low-quality samples were excluded according to the same criteria adopted in study \cite{zhou2023assessing}. Samples with mapping rates below 80\% or fewer than 10,000 detected genes were removed. The remaining gene counts were normalised using the variance-stabilising transformation implemented in DESeq2 (for ML methods), and gene identifiers were harmonised to Ensembl symbols. Only proteincoding genes consistently expressed across all samples were retained to ensure comparability. In this study, we use the available preprocessing data and matrices from \cite{zhou2023assessing}. Each record comprised approximately 17,000 gene features and associated metadata, including biomaterial type, lineage label, and induction time point. Table \ref{tab:dataset_composition} summarises the dataset composition used for model training and evaluation.

\begin{table}[htbp]
\centering
\caption{Summary of dataset composition used for training and evaluation.}
\label{tab:dataset_composition}
\begin{tabular}{lccc}
\hline
\textbf{Lineage Type} & \textbf{Training Samples} & \textbf{Test Samples} & \textbf{Total} \\
\hline
Osteogenic        & 40  & 23 & 63  \\
Chondrogenic      & 26  & 11 & 37  \\
Adipogenic        & 39  & 12 & 51  \\
Undifferentiated  & 19  & 20 & 39  \\
\hline
Total             & 124 & 66 & 190 \\
\hline
\end{tabular}
\end{table}

\subsection{Experimental Settings}
Following the training pipeline introduced in Section 3.2,
the proposed framework was trained in two stages. In the first stage, PuritySCVI was trained in an unsupervised manner to learn the latent structure and infer sample purity. The latent dimension was set to 10, the learning rate was fixed at $1\times10^{-3}$, and training was run for up to 500 epochs. During the early warm-up period, the GBMPurity head was frozen and later gradually unfrozen for joint optimisation. In the second stage, PuritySCANVI was initialized from the pretrained PuritySCVI and further fine-tuned in a semi-supervised manner with lineage labels. The learning rate was set to $5\times10^{-4}$, and early stopping with a patience of 100 epochs was adopted to avoid overfitting. To ensure reproducibility and evaluate training stability, all experiments were repeated under six random seeds: 1, 7, 77, 89, 777, and 999. 

A curriculum strategy is adopted throughout training. At the beginning, the model places stronger emphasis on smoothness and purity-related regularisation. As training progresses, these weights are gradually reduced, allowing the classifier to focus more on stable lineage discrimination.

Model performance was then evaluated on the held-out test set using classification accuracy and weighted F1-score. In addition, confusion matrices were analysed to examine class-level prediction errors, and temporal purity trajectories were inspected to assess the biological consistency of the learned representations.

We compare our proposed framework with (i) Traditional machine-learning baselines (k-nearest neighbours and random forests) from the MeD-P framework. (ii) Standard scVI and scANVI models trained without purity estimation.

\subsection{Experimental Results}

\begin{figure*}
    \centering
    \includegraphics[width=1.0\linewidth]{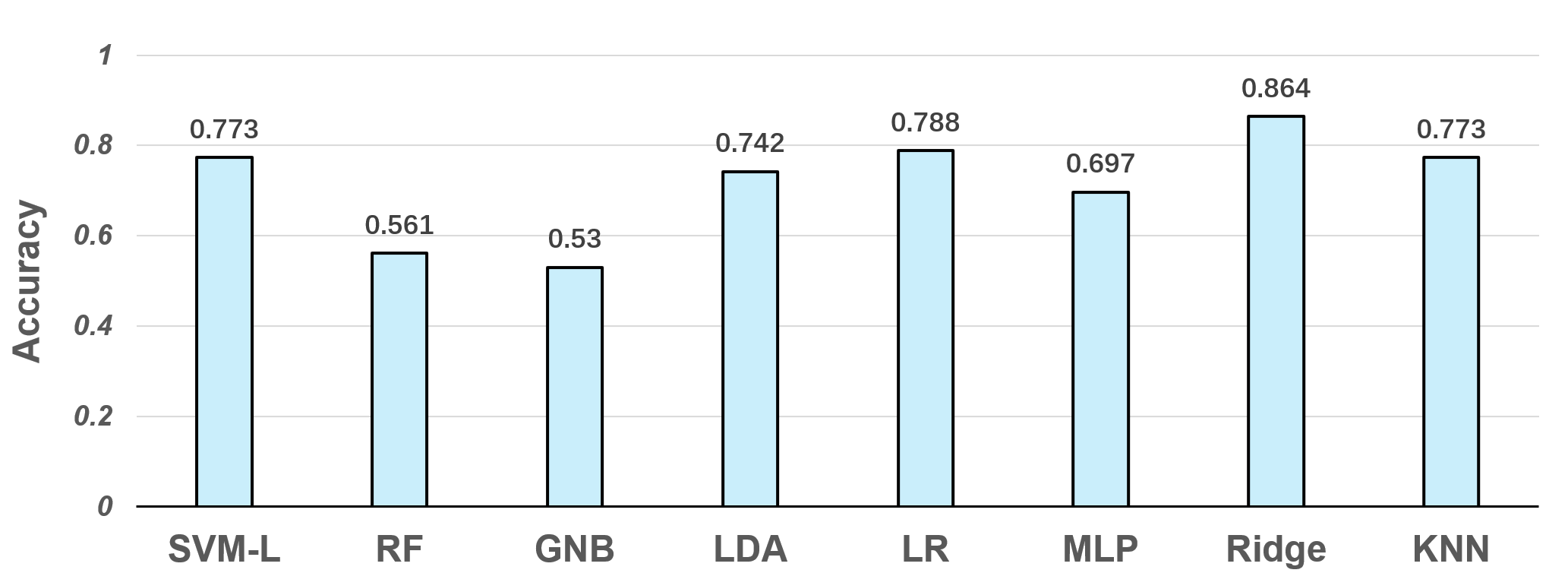}
    \caption{Prediction accuracy of traditional MeD-P classifiers.}
    \label{fig:Med}
\end{figure*}

Figure \ref{fig:Med} and Table \ref{tab:table42} illustrate the prediction accuracies obtained from the  MeD-P framework and the proposed purity-aware deep generative models. In the MeD-P experiment, traditional classifiers such as support vector machines and random forests achieved accuracies ranging from 0.530 to 0.864, with Ridge achieving the best performance among the reproduced traditional baselines. When replicated using the dataset reprocessed for this study, variations in sample composition and normalisation slightly altered model performance. By contrast, the proposed PuritySCVI and PuritySCANVI models achieved the highest mean accuracy of 0.932 ± 0.009, surpassing both traditional methods and the standard scVI with scANVI configuration (0.921 ± 0.019). These findings demonstrate that incorporating purity estimation and regularisation enhances both accuracy and generalisation across random initializations.

Table \ref{tab:table42} also  reports the detailed results for each random seed, including both accuracy and weighted F1-scores. The purity-aware models shows a better performance the baseline scVI with scANVI pipeline in mean accuracy and exhibited lower variance, indicating improved stability across runs. These improvements arise from purity-guided loss components and pretrained compositional priors that stabilise latent-space optimisation and enhance lineage discriminability. The proposed framework achieved a mean accuracy of 0.932 ± 0.009 and a mean weighted F1-score of 0.934 ± 0.008, exceeding the baseline averages of 0.921 ± 0.019 and 0.923 ± 0.020, respectively. The smaller standard deviations reflect better consistency across different initialisations, confirming the robustness of the proposed purity-aware architecture in modelling lineage commitment under biomaterial conditions.

\begin{table*}[t]
\centering 
\caption{Performance comparison between the proposed PuriGen and the standard scVI with scANVI baseline across six random seeds.}
\label{tab:table42}
\setlength{\tabcolsep}{6pt}
\renewcommand{\arraystretch}{1.1}
\begin{tabular*}{\textwidth}{@{\extracolsep{\fill}}c|cc|cc@{}}
\hline
& \multicolumn{2}{c|}{\textbf{Accuracy}}  
& \multicolumn{2}{c}{\textbf{F1 Score}} \\
\hline
\textbf{Seed} 
& PuriGen (Ours) 
& scVI+scANVI 
& PuriGen (Ours)
& scVI+scANVI \\
\hline
1   & 0.924 & 0.924 & 0.928 & 0.928 \\
7   & 0.939 & 0.924 & 0.941 & 0.928 \\
77  & 0.924 & 0.894 & 0.926 & 0.894 \\
89  & 0.940 & 0.940 & 0.941 & 0.942 \\
777 & 0.924 & 0.894 & 0.928 & 0.898 \\
999 & 0.940 & 0.940 & 0.941 & 0.941 \\
\hline
\textbf{Mean} 
& 0.932 ($\pm$ 0.009) 
& 0.921 ($\pm$ 0.019) 
& 0.934 ($\pm$ 0.008) 
& 0.923 ($\pm$ 0.020) \\
\hline
\end{tabular*}
\end{table*}

Furthermore, model predictions for the test subset were generated to visualize classification behaviour.
The confusion matrix shown in Figure 3 highlights class-level performance. Most misclassifications were biased toward the adipogenic class, particularly for osteogenic samples, whereas chondrogenic samples were correctly identified in this test split. The PuritySCANVI embeddings showed better class separation and smoother temporal progression than the baseline scANVI model. The addition of purity information enabled the model to capture continuous differentiation trajectories rather than discrete class boundaries. This improvement indicates that purity regularisation helped align the latent space with biological development stages.\\

\begin{figure}
    \centering
    \includegraphics[width=0.8\linewidth]{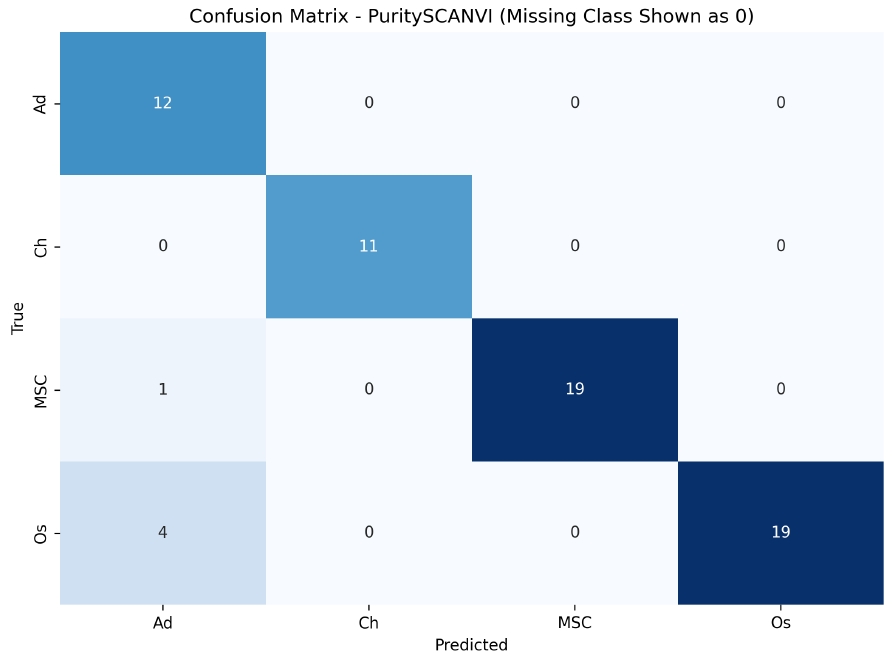}
    \caption{Confusion matrix of the proposed PuriGen prediction performance.}
    \label{fig:placeholder}
\end{figure}

\textit{Summary: }The results confirm that integrating purity estimation into a probabilistic generative framework improves both prediction accuracy and interpretability. The proposed framework achieved higher accuracy and F1 scores than baseline methods, with lower variance across runs. The purity-aware design enhanced latentspace structure, providing a clearer separation of lineages and smoother temporal trends. Compared with traditional discriminative models, the approach not only predicts lineage fate but also quantifies sample composition, adding interpretive depth to bulk transcriptomic analysis.

These experiments validate the effectiveness of purity-aware generative modelling for biomaterial-induced stem-cell lineage prediction. The unified training and evaluation pipeline demonstrates that combining pretrained purity priors with latent-variable inference yields a stable, interpretable, and biologically coherent model.

\section{Conclusion}

This work proposed a purity-aware deep generative framework (PuriGen) for predicting biomaterial-induced stem-cell lineage fate from bulk RNA-seq data. By informing scVI and scANVI with purity estimation, the proposed PuritySCVI and PuritySCANVI models jointly learned latent representations, lineage labels, and purity-related information within a unified probabilistic framework. In addition, purity-aware and temporal regularisation terms were introduced to encourage biologically meaningful latent structures and stable optimisation. Experimental results showed that the PuriGen achieves a higher mean accuracy and weighted F1-score, with lower variance across six random initializations. Beyond predictive performance, the learned latent space and purity trajectories provided improved interpretability, suggesting that explicitly modelling compositional heterogeneity can better capture differentiation dynamics in bulk transcriptomic data. 
Future work includes exploring adaptive purity modelling, multimodal extensions, and validation on additional datasets and biological systems.
%
%
%
\bibliographystyle{splncs04} 
\bibliography{mybibliography}

\end{document}